\documentclass{ivs2e}     

\PaperSize{A4}

\Title{Opacity in the jet of 3C\,309.1}

\STitle{Opacity in the jet of 3C\,309.1}

\PublicationType{proceedings} 

\PublicationName{15th European VLBI Meeting Proceedings}

\IvsComponentName{Coordinating Center}

\NumberofAuthors{2}

\AuthorName{1}{Eduardo Ros}
\AuthorName{2}{Andrei P. Lobanov}

\ContactAuthorReference{1}

\AuthorEmail{1}{ros@mpifr-bonn.mpg.de}
\AuthorEmail{2}{alobanov@mpifr-bonn.mpg.de}

\AuthorTelephone{1}{+49-228-525292}
\AuthorTelephone{2}{+49-228-525295}

\AuthorPersonalWebPage{1}{http://www.mpifr-bonn.mpg.de/staff/ros}   
\AuthorPersonalWebPage{2}{http://www.mpifr-bonn.mpg.de/staff/alobanov}   

\AuthorInstitutionReference{1}{1}
\AuthorInstitutionReference{2}{1}

\NumberofInstitutions{1}

\InstitutionName{1}{Max-Planck-Institut f\"ur Radioastronomie}

\InstitutionPostAddress{1}{Auf dem H\"ugel 69, D-53121 Bonn}

\InstitutionCountry{1}{Germany}

\InstitutionFax{1}{+49-228-525229}

\InstitutionWebPage{1}{http://www.mpifr-bonn.mpg.de}
\newcommand\plot[1]{
\includegraphics{#1.ps}
}

\newcommand{\mton}[1]{ \multicolumn{1}{c}{ #1 } }
\newcommand{\mtwc}[1]{ \multicolumn{2}{c}{ #1 } }
\newcommand{\mthr}[1]{ \multicolumn{3}{c}{ #1 } }

\newcommand{\mtsx}[1]{ \multicolumn{6}{c}{ #1 } }

\begin{document}

%
%
\MakeTitle                

%
%
\begin{abstract}
The ``core" of a radio source is believed to mark the
frequency-dependent location where the optical depth to 
synchrotron self--absorption $\tau_s \approx 1$.  The frequency
dependence can be used for derive physical conditions of the 
radio emitting region and the ambient environment near the central engine
of the radio source. 
In order to test and improve the models to derive this
information, we made multi-frequency 
dual-polarization observations of 3C\,309.1 in 1998.6,
phase-referenced to the QSOs
S5\,1448+76 and 4C\,72.20 (S5\,1520+72) ($4^\circ26^\prime$ and
$1^\circ49^\prime$ away, respectively).
We present here preliminary results from these observations:
total and polarized intensity maps, spectral information
deduced from these images, and the relative position
of 3C\,309.1 with respect to S5\,1448+76 at different
frequencies.  Finally, we discuss briefly the observed shift 
of the core position.  
\end{abstract}

%
\section{Introduction\label{sec:introduction}}

\paragraph{Opacity in pc-scale jets}
The unresolved ``core" of a compact extragalactic
radio source is believed to mark the
location where the optical depth to synchrotron self absorption
$\tau_s \approx 1$.  This position
changes with observing frequency as 
$R_{\rm core} \propto \nu^{-1/k_r}$ 
\cite{koe81}.  The power index $k_r$ 
depends on the shape of the
electron energy spectrum and on the magnetic field and particle density 
distributions in the ultra-compact jet.
Hence, by studying variations at
$k_r$ as a function of frequency, we may study the detailed physical
conditions of the radio emitting region and the ambient environment of the
source very near the central engine.  
Following \cite{lob98} we can estimate 
basic physical parameters of the jet including
luminosity, maximum
brightness temperature, magnetic field in the jet, particle density,
and the geometrical properties of the jet (i.e., the core location
respect to the jet origin).

To measure $k_r$ a knowledge of the absolute position of the core 
(or at least the core offset between different frequencies) is needed.
Hybrid maps in VLBI lack this positional information due to the use of 
closure-phase in the imaging process.
The rigorous alignment
of hybrid maps can be made by astrometric phase-referencing
(e.g.\ \cite{leb01,bri00,per00,ros99}).  
In sources with extended structure, an optically thin component can 
be used to align maps at different frequencies and then estimate 
the position of the core.  

\paragraph{The QSO 3C\,309.1}
The QSO {3C\,309.1} 
(V$=$16.78, $z$=0.905\footnote{This redshift corresponds to a linear scale
of 5.60\,$h^{-1}$\,pc\,mas$^{-1}$
for $H_0$=75\,$h$\,km\,s$^{-1}$\,Mpc$^{-1}$ and $q_0$=0.5.})
is one of the most prominent compact steep spectrum (CSS) radio
sources \cite{vbr84,fan85}.
Many of the CSS sources display polarized emission at cm-wavelengths.
The ionized gas surrounding the jet disrupts
its flow and is responsible of the complexity of the radio structures
seen.  
There is evidence suggesting that 3C\,309.1 is located at
the center of a very massive cooling flow with $\dot{M}>1000\,{\rm M}_\odot\,
{\rm yr}^{-1}$ within a radius of 11.5\,h$^{-1}$\,kpc
\cite{for90}.

VLBA observations of 3C\,309.1 at 6 frequencies were used for determining the 
behavior of $k_r$ between 1.6 and 22\,GHz \cite{lob98}.  To improve
and extend this determination to higher frequencies, we observed 3C\,309.1
using the VLBA at eight
frequencies with dual polarization.  In this contribution,
we present a preliminary analysis of the new observations.

\section{Mapping analysis\label{sec:results}}

\paragraph{VLBA Observations\label{par:observations}}
We carried out VLBA multi-frequency, dual-polarization observations of 
{3C\,309.1} on July 19th and 23rd 1998 
using all 10 VLBA antennas and observing at
1.5, 1.6, 2.3, 5, 8.4, 15, 22, and 43\,GHz.
The data were correlated at the
NRAO\footnote{VLBA correlator, Array Operations
Center, National Radio Astronomy Observatory (NRAO), Socorro, NM.}
and processed using {\sc aips}\footnote{Astronomical 
Image Processing System, developed and maintained by the NRAO.} and
{\sc difmap} \cite{she95}.  {\sc difmap} was used for
mapping the total intensity emission.
The polarized intensity mapping and the phase-referencing
analysis were carried out in {\sc aips}.
The 43\,GHz data had to be discarded due to calibration
and coherence problems.

\begin{figure}[htbp]
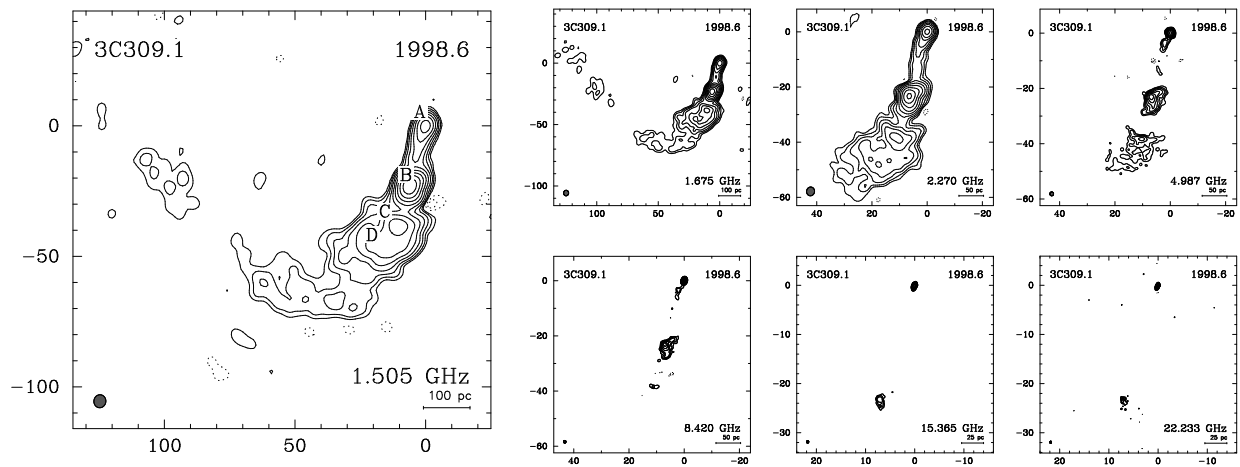

\vspace{60pt}
\begin{tabular}{p{80pt}p{80pt}p{80pt}p{80pt}p{80pt}}
\includegraphics{ros101.ps} & &
\includegraphics{ros102.ps} &
\includegraphics{ros103.ps} &
\includegraphics{ros104.ps} \\[80pt]
& & \includegraphics{ros105.ps} &
\includegraphics{ros106.ps}   &
\includegraphics{ros107.ps}  \\[30pt]
\end{tabular}
\caption{VLBA total intensity
images of 3C\,309.1. 
The synthesized interferometric beams
are represented at the bottom left corner of each image.  
Contour levels are drawn at
$\sqrt{2}$ intervals.
The left image follows the labelling convention from \cite{kus90}.
Map parameters (beam, total flux density, peak of brightness, lowest contour
in map) are given in Table \ref{table:mapparameters}.
\label{fig:3c309maps}
}
\end{figure}

\paragraph{Total intensity mapping\label{subsec:Imaps}}
We applied the {\sc clean} algorithm and self-calibration in
{\sc difmap} to 
obtain the total intensity maps presented in Fig.\ \ref{fig:3c309maps}
and described in Table~\ref{table:mapparameters}.
The resulting images show a structure very similar to the reported in
earlier works (see
\cite{aar96} and references therein).
The source shows a core-jet structure first oriented southward and turning
later to the East at about 60\,mas from the core.

%

%
\begin{table}[htbp]
\caption{Total intensity map parameters 
(Figs.\ \ref{fig:3c309maps} \& \ref{fig:maps1448and1520})}
\label{table:mapparameters}
\begin{scriptsize}
\begin{center}
\begin{tabular}{@{}r@{\,\,}
                r@{$\times$}l@{\,\,}r@{\,\,}c@{\,\,}c@{\,\,}c@{\,\,}
                r@{$\times$}l@{\,\,}r@{\,\,}c@{\,\,}c@{\,\,}c@{\,\,}
                r@{$\times$}l@{\,\,}r@{\,\,}c@{\,\,}c@{\,\,}c@{}
}
      & \mtsx{\bf --- 3C\,309.1 ---}
      & \mtsx{\bf --- 4C\,72.20 ---}
      & \mtsx{\bf --- S5\,1448+76 ---} \\
      & \mthr{Beam}            &                 &      & &
       \mthr{Beam}            &                 &      & &
       \mthr{Beam}            &                 &      & \\
$\nu$ & \mtwc{size}    & P.A.\ & $S_{\rm peak}$  & $S_{\rm min}$$^{\rm (a)}$ & $S_{\rm tot}$$^{\rm (b)}$ 
                                                          & \mtwc{size}  & P.A.\ & $S_{\rm peak}$ & $S_{\rm min}$$^{\rm (a)}$  & $S_{\rm tot}$$^{\rm (b)}$ 
                                                                                                  & \mtwc{size}   & P.A.\ & $S_{\rm peak}$ & $S_{\rm min}$$^{\rm (a)}$ & $S_{\rm tot}$$^{\rm (b)}$ \\	
{\tiny [Hz]}  & \mtwc{\tiny [mas]}   
                       & {\tiny [$^\circ$]} 	
                               & {\tiny [Jy/b]}  & {\tiny [mJy/b]}       
                                                & {\tiny [Jy]}  
                                                           & \mtwc{\tiny [mas]}   
                                                                          & {\tiny [$^\circ$]} 	
                                                                                  & {\tiny [Jy/b]}        & {\tiny [mJy/b]}       
                                                                                          & {\tiny [Jy]} 
                                                                                                  & \mtwc{\tiny [mas]}   
                                                                                                                  & {\tiny [$^\circ$]} 	
                                                                                                                          & {\tiny [Jy/b]} & {\tiny [mJy/b]}       
                                                                                                                                   & {\tiny [Jy]}  \\ \hline
1.505 & 5.00  & 4.60 & --7.5 & 0.972$^{\rm c}$ & 2.0 & 3.424 & 6.47 & 5.96 & --5.5 & 0.053 & 0.5 & 0.055 & 6.49 & 6.21 &   1.7 & 0.145 & 0.6 & 0.171 \\
1.675 & 4.50  & 4.10 & --4.3 & 0.819$^{\rm c}$ & 2.0 & 3.181 & 5.83 & 5.34 & --4.5 & 0.050 & 0.4 & 0.052 & 5.77 & 5.50 & --3.3 & 0.149 & 0.6 & 0.174 \\
2.270 & 3.19  & 2.90 &--10.5 & 0.604$^{\rm c}$ & 2.0 & 2.699 & 4.09 & 3.68 & --8.4 & 0.083 & 0.8 & 0.085 & 4.11 & 3.74 &--13.8 & 0.225 & 0.8 & 0.253 \\
4.987 & 1.40  & 1.35 &   8.5 & 0.675$^{\rm d}$ & 2.0 & 1.797 & 1.90 & 1.82 &--13.6 & 0.053 & 0.4 & 0.065 & 1.88 & 1.79 &--24.8 & 0.263 & 0.6 & 0.313 \\
8.420 & 0.95  & 0.85 & --5.0 & 0.646$^{\rm d}$ & 2.0 & 1.326 & 1.08 & 0.99 & --6.7 & 0.056 & 0.8 & 0.057 & 1.10 & 1.02 &--11.0 & 0.216 & 0.8 & 0.303 \\
15.365& 0.47  & 0.45 &--17.8 & 0.492$^{\rm d}$ & 3.0 & 0.921 & 0.58 & 0.56 &--14.8 & 0.100 & 0.8 & 0.101 & 0.59 & 0.58 &   4.7 & 0.156 & 1.1 & 0.244 \\
22.233& 0.41  & 0.33 &--14.8 & 0.430$^{\rm d}$ & 3.0 & 0.712 & 0.54 & 0.45 & --0.2 & 0.195 & 1.5 & 0.205 & 0.55 & 0.45 & --6.6 & 0.211 & 1.6 & 0.241 \\ \hline
\end{tabular}
\end{center}
$^{\rm a}$ Minimum contour level in the figure.
$^{\rm b}$ Total flux density recovered in the map model.  
$^{\rm c}$ Corresponds to the B component.
$^{\rm d}$ Corresponds to the A component.
\end{scriptsize}
\end{table}

\begin{figure}[htbp]
%
\vspace{203pt}
\includegraphics{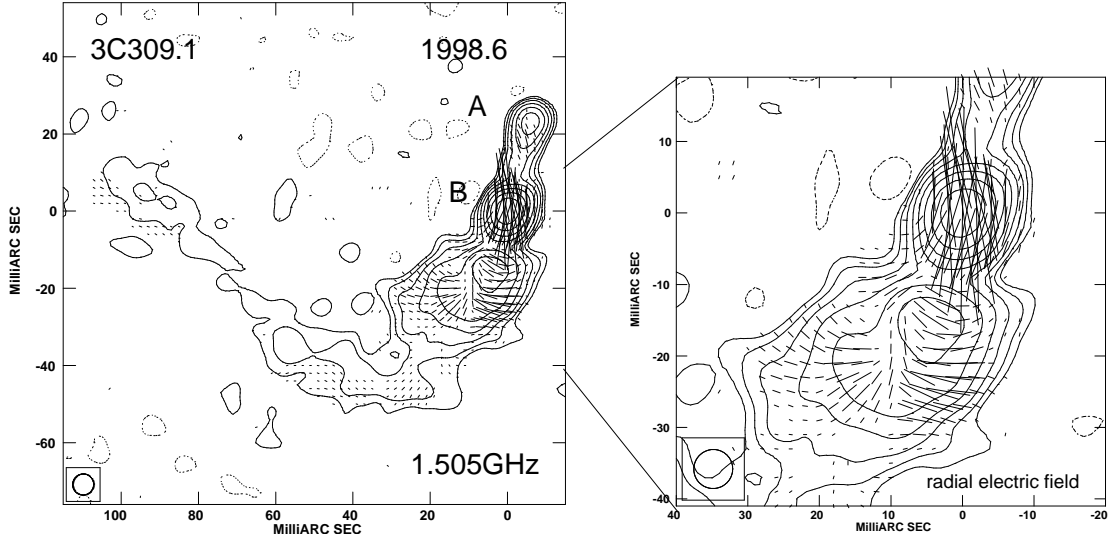}
\caption{Polarized intensity image of 
3C\,309.1 at 1.505\,GHz ($\lambda$\,21cm). 
Electric field vectors are shown.
\label{fig:mappol}
}
\end{figure}

\paragraph{Polarized intensity maps\label{subsec:Pmaps}}
We applied the instrumental polarization calibration 
from the total intensity maps using the {\sc aips} task {\sc lpcal}
as described in \cite{lep95}.  We imaged the Stokes Q and U and produced
images of the linearly polarized emission and the electric vector
position angle.  An exhaustive description of these results will
be published elsewhere.  We show an image of the linear polarization
distribution at 1.5\,GHz in Fig.~\ref{fig:mappol}.  The core is unpolarized
as in many QSOs.  In the region South of B the electric
vector is radial, suggesting a toroidal magnetic field viewed edge-on.
The degree of polarization is higher at the outer parts of the jet.

%

\paragraph{Spectral analysis\label{subsec:spectra}}
The overall spectral index of {3C309.1} is 
$-0.57\pm0.01$.  This result is obtained by adding together
emission from
all of the radio source structure which may have very different
physical properties.

\begin{figure}[htbp]
%
\vspace{280pt}
\includegraphics{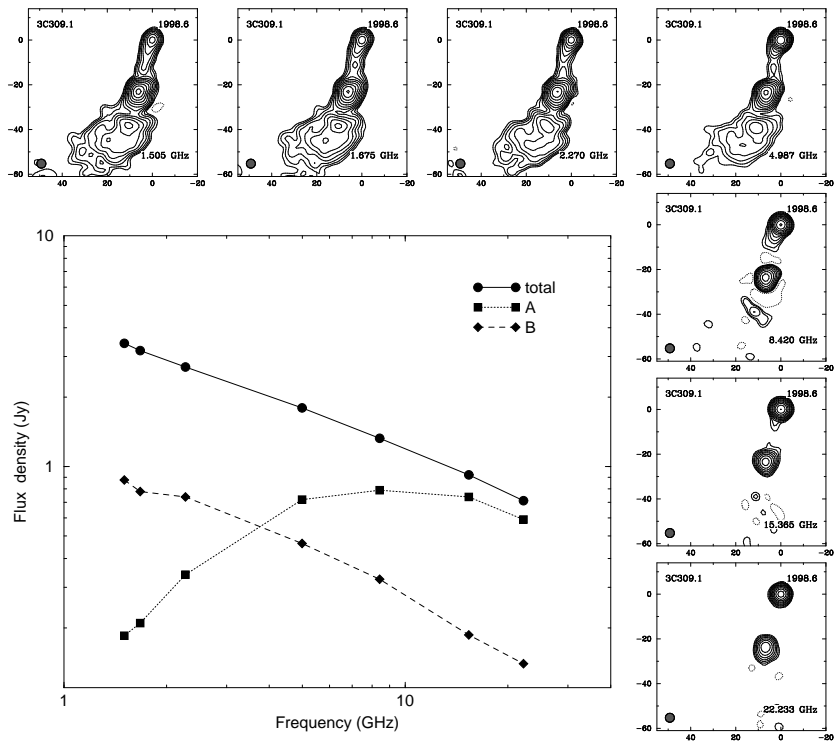}
\caption{Maps of 3C\,309.1 
at 4\,mas resolution, obtained using the natural weighting 
and Gaussian tapering with the half maximum at 33\,M$\lambda$.  The
contour levels are drawn at $\sqrt{3}$ intervals.  The lowest level is
of 3\,mJy/beam}
\label{fig:allmaps4}
\end{figure}

To study spectral properties at different parts of the source,
we mapped the radio source at
all frequencies using natural weighting and
very strong tapering (Gaussian
function with half maximum at a distance of 33\,M$\lambda$). 
We convolved the {\sc clean} components with a circular beam of 
4\,mas in size, aligning the images on the peak-of-brightness 
of component A.  We show these images in
Fig.\ \ref{fig:allmaps4} together with the spectra of
components A and B
and the total spectrum of the VLBI emission.
The turnover frequency for A is around
8.4\,GHz, and is below 1.4\,GHz for B.  A linear regression to
the points for B provides an overall spectral index of $-0.67\pm0.04$.

%

\section{Phase-referencing analysis}

\begin{figure}[htbp]
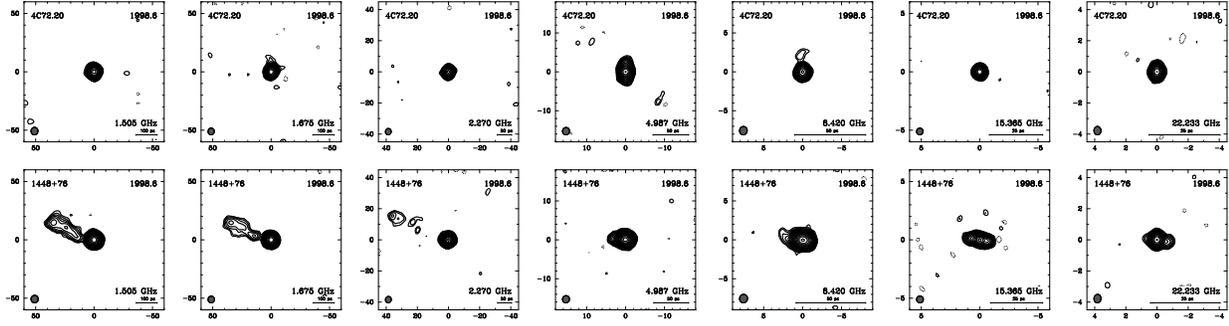

\vspace{50pt}
\begin{tabular}{p{55pt}p{55pt}p{55pt}p{55pt}p{55pt}p{55pt}p{55pt}}
\plot{ros401} &
\plot{ros402} &
\plot{ros403} &
\plot{ros404} &
\plot{ros405} &
\plot{ros406}  &
\plot{ros407} \\[50pt]
\plot{ros408} &
\plot{ros409} &
\plot{ros410} &
\plot{ros411} &
\plot{ros412} &
\plot{ros413} &
\plot{ros414} \\
\end{tabular}
\caption{Hybrid maps of the position
calibrator sources 4C\,72.20 (top) and S5\,1448+76 (bottom).
The map parameters are
described in Table \ref{table:mapparameters}.
\label{fig:maps1448and1520}
}
\end{figure}

\paragraph{The calibrators}
We used two position calibrators for 3C\,309.1:
\begin{itemize}
\item 4C\,72.20 is a QSO with $z$=$0.799$ 
and $V$=$16.5$, $1^\circ49^\prime$ 
East of the target source.  It is a point-like source with an inverted
spectrum.  The imaging results are presented in 
Fig.~\ref{fig:maps1448and1520}.  
\item S5\,1448+76 is a flat spectrum, compact radio source with $z$=$0.899$
and $V$=$20.0$.  It is $4^\circ26^\prime$ to the NW of 3C\,309.1.   
It shows a faint jet to the NE at the lower frequencies, and
it is also elongated in the East-West direction at the higher frequencies.
The hybrid maps are shown in Fig.~\ref{fig:maps1448and1520}.  
\end{itemize}

\paragraph{The analysis}
We carried out the phase-referencing analysis in {\sc aips}.  
We solved for the phase, delay and phase-rate for 3C\,309.1, using the
total intensity maps as input (dividing the $(u,v)$-data by the 
{\sc clean} model) and thus 
removing the effect of the source structure.
We then interpolated the values fitted using the task
{\sc clcal} for the fainter phase-reference calibrators, S5\,1442+76 and
4C\,72.20 (S5\,1520+76). 

\begin{figure}[htbp]
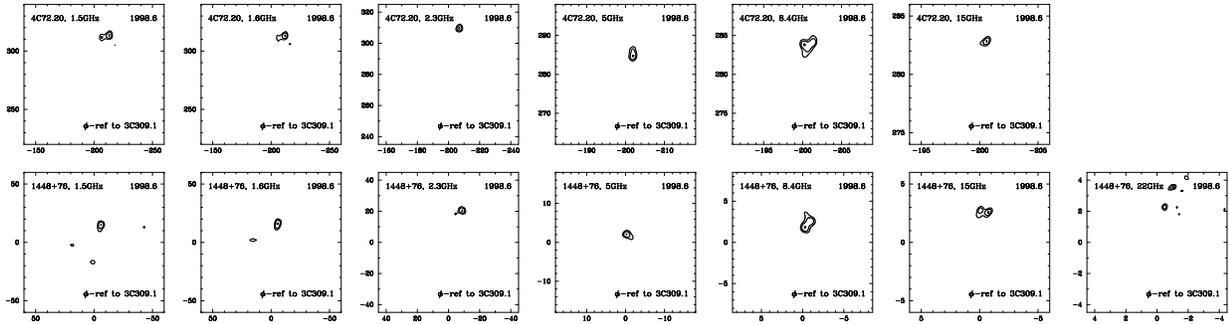

\vspace{50pt}
\begin{tabular}{p{55pt}p{55pt}p{55pt}p{55pt}p{55pt}p{55pt}p{55pt}}
\plot{ros501} &
\plot{ros502} &
\plot{ros503} &
\plot{ros504} &
\plot{ros505} &
\plot{ros506}  &
\\[50pt]
\plot{ros507} &
\plot{ros508} &
\plot{ros509} &
\plot{ros510} &
\plot{ros511} &
\plot{ros512} &
\plot{ros513} \\
\end{tabular}
\caption{Phase-reference maps of 4C\,72.20 and S5\,1448+76 obtained
using the phase, delay and phase-rate solutions for 3C\,309.1 as described
in the text.  The contours are 49, 69 and 98\% of the peak of brightness
for each map.  The values of the brightness peaks 
and their ratio (in percentage) with respect
to their  peaks in the hybrid 
maps (Fig.\ \ref{fig:maps1448and1520})
are given in
Table~\ref{table:maprefparameters}.
\label{fig:mapsref1448and1520}
}
\end{figure}

After editing the data, we mapped the radio sources using the 
{\sc aips} task {\sc imagr} with the same parameters as were used for
the hybrid imaging in {\sc difmap}.
The phase-referenced maps are shown in 
Fig.~\ref{fig:mapsref1448and1520} and the corresponding map parameters 
are given in
Table~\ref{table:maprefparameters}.
We measured the positions of the
brightness peaks, whose offsets from the coordinate
origin correspond to the offsets 
from the nominal position of 3C\,309.1 relative to the calibrators.
The relative positions deduced from this procedure are presented
in Table~\ref{table:positions}.

\begin{table}[htbp]
\caption{Phase-referenced map parameters for 4C\,72.20 and S5\,1448+76 (Fig.\
\ref{fig:mapsref1448and1520})
\label{table:maprefparameters}
}
\begin{footnotesize}
\begin{center}
\begin{tabular}{rrrrrrr}
 & \mthr{\bf --- 4C\,72.20 ---} & \mthr{\bf --- S5\,1448+76 ---} \\
$\nu$ &
\mton{$S_{\rm max}^{\phi{\rm -ref}}$} &
\mton{$S_{\rm max}^{\rm hybrid}$} &
\mton{$\frac{S_{\rm max}^{\phi{\rm -ref}}}{S_{\rm max}^{\rm hybrid}}$} &
\mton{$S_{\rm max}^{\phi{\rm -ref}}$} &
\mton{$S_{\rm max}^{\rm hybrid}$} &
\mton{$\frac{S_{\rm max}^{\phi{\rm -ref}}}{S_{\rm max}^{\rm hybrid}}$} \\
\,[GHz] & \mton{[{\tiny mJy/beam}]} & \mton{[{\tiny mJy/beam}]} & [\%] & 
\mton{[{\tiny mJy/beam}]} & \mton{[{\tiny mJy/beam}]} & [\%] \\ \hline
 1.505 & 13.6 &  53.3 & 25.5	& 29.2 & 145.2 & 20.1 \\
 1.675 & 14.8 &  49.7 & 29.8	& 33.3 & 148.7 & 22.4 \\
 2.270 & 30.0 &  82.3 & 36.5	& 82.4 & 225.3 & 36.6 \\
 4.987 & 19.0 &  52.8 & 36.0	&143.7 & 263.2 & 54.6 \\
 8.420 &  7.1 &  55.3 & 12.8	& 55.3 & 216.3 & 25.6 \\
15.365 &  4.7 & 100.4 &  4.7	& 22.2 & 155.8 & 14.2 \\
22.233 &  \mthr{---}    	& 11.9 & 211.3 &  5.6 \\ \hline
\end{tabular}
\end{center}
\end{footnotesize}
\end{table}

\begin{table}[htbp]
\caption{Relative right ascensions and declinations in J2000.0 coordinates 
of 3C\,309.1  with respect to 4C\,72.20, and S5\,1448+76, obtained via
{\sc aips} phase-referencing.  Biases have not been corrected.
\label{table:positions}}
\begin{footnotesize}
\begin{center}
\begin{tabular}{
r
r@{$^{\rm h}$}
r@{$^{\rm m}$}
r@{$\rlap{.}^{\rm s}$}
l@{$\pm$}
r@{$\rlap{.}^{\rm s}$}
l
r@{$^\circ$}
r@{$'$}
r@{$\rlap{.}''$}
l@{$\pm$}
r@{$\rlap{.}''$}
l
r@{$^{\rm h}$}
r@{$^{\rm m}$}
r@{$\rlap{.}^{\rm s}$}
l@{$\pm$}
r@{$\rlap{.}^{\rm s}$}
l
r@{$^\circ$}
r@{$'$}
r@{$\rlap{.}''$}
l@{$\pm$}
r@{$\rlap{.}''$}
l
c
}
& \multicolumn{12}{c}{\bf --- 4C\,72.20 ---} &
 \multicolumn{12}{c}{\bf --- S5\,1448+76 ---} \\
$\nu$ [GHz] 
& \multicolumn{6}{c}{$\Delta \alpha_{({\rm 3C\,309.1~-~4C\,72.20})}$}
& \multicolumn{6}{c}{$\Delta \delta_{({\rm 3C\,309.1~-~4C\,72.20})}$} 
& \multicolumn{6}{c}{$\Delta \alpha_{({\rm 3C\,309.1~-~S5\,1448+76})}$}
& \multicolumn{6}{c}{$\Delta \delta_{({\rm 3C\,309.1~-~S5\,1448+76})}$} \\ \hline
 1.505 
& --0 & 21 & 40 & 0554 & 0 & 0015 & --0 & 44 & 45 & 712 & 0 & 007 
& 0 & 10 & 38 & 805   & 0 & 002   & --4 & 20 & 51 & 721  & 0 & 006 \\ 
 1.675 
& --0 & 21 & 40 & 0557 & 0 & 0015 & --0 & 44 & 45 & 712 & 0 & 007 
& 0 & 10 & 38 & 8045  & 0 & 0014  & --4 & 20 & 51 & 722  & 0 & 005 \\
 2.270 
& --0 & 21 & 40 & 0568 & 0 & 0014 & --0 & 44 & 45 & 708 & 0 & 006 
& 0 & 10 & 38 & 8053  & 0 & 0007  & --4 & 20 & 51 & 726  & 0 & 003 \\
 4.987 
& --0 & 21 & 40 & 0565 & 0 & 0014 & --0 & 44 & 45 & 707 & 0 & 006 
& 0 & 10 & 30 & 80476 & 0 & 00017 & --4 & 20 & 51 & 7313 & 0 & 0006 \\
 8.420 
& --0 & 21 & 40 & 0569 & 0 & 0014 & --0 & 44 & 45 & 706 & 0 & 006 
& 0 & 10 & 30 & 80482 & 0 & 00009 & --4 & 20 & 51 & 7320 & 0 & 0003 \\
15.365 
& --0 & 21 & 40 & 0568 & 0 & 0014 & --0 & 44 & 45 & 705 & 0 & 006 
& 0 & 10 & 30 & 80489 & 0 & 00008 & --4 & 20 & 51 & 7318 & 0 & 0003 \\
22.233 
& \mtsx{---$^{\rm (a)}$} & \mtsx{---$^{\rm (a)}$} 
& 0 & 10 & 30 & 80482 & 0 & 00007 & --4 & 20 & 51 & 7315 & 0 & 0003 \\ \hline
\end{tabular}
\end{center}
$^{\rm a}$ No phase-referencing detection.
\end{footnotesize}
\end{table}

The catalogue position of 4C\,72.20
used at the correlator in error of +200 in $\alpha$ and --280\,mas in $\delta$.
This translates into an
estimated uncertainty of $\sim$6\,mas, in our preliminary 
position determination at each frequency, making this
fraction of the data unusable for our purposes.
A proper analysis, correcting for the wrong position of 4C\,72.20 
will be published elsewhere.
No ionosphere corrections
have been applied in the data analysis.  At frequencies
lower than 8.4\,GHz, the ionospheric dispersion may severely bias
our results.
The tropospheric delay, especially the wet part, affects the phase
for the highest frequencies, where the size of the water particles
in the atmosphere is comparable to the wavelength.  

Notice that the ratio between the peaks of
brightness of the phase-referenced maps and the hybrid maps
(4$^{\rm th}$ and 7$^{\rm th}$ columns in Table~\ref{table:maprefparameters})
is the highest at the intermediate frequencies, where the compromise
between the ionospheric and the tropospheric effects is found.  Even
when the {\em a priori} position of 4C\,72.20 is in error, its 
ratios are similar to the ones in S5\,1448+76, probably because the former 
is $\sim$3 times closer to 3C\,309.1 than the latter.

\begin{figure}[htbp]
\vspace{150pt}
\includegraphics{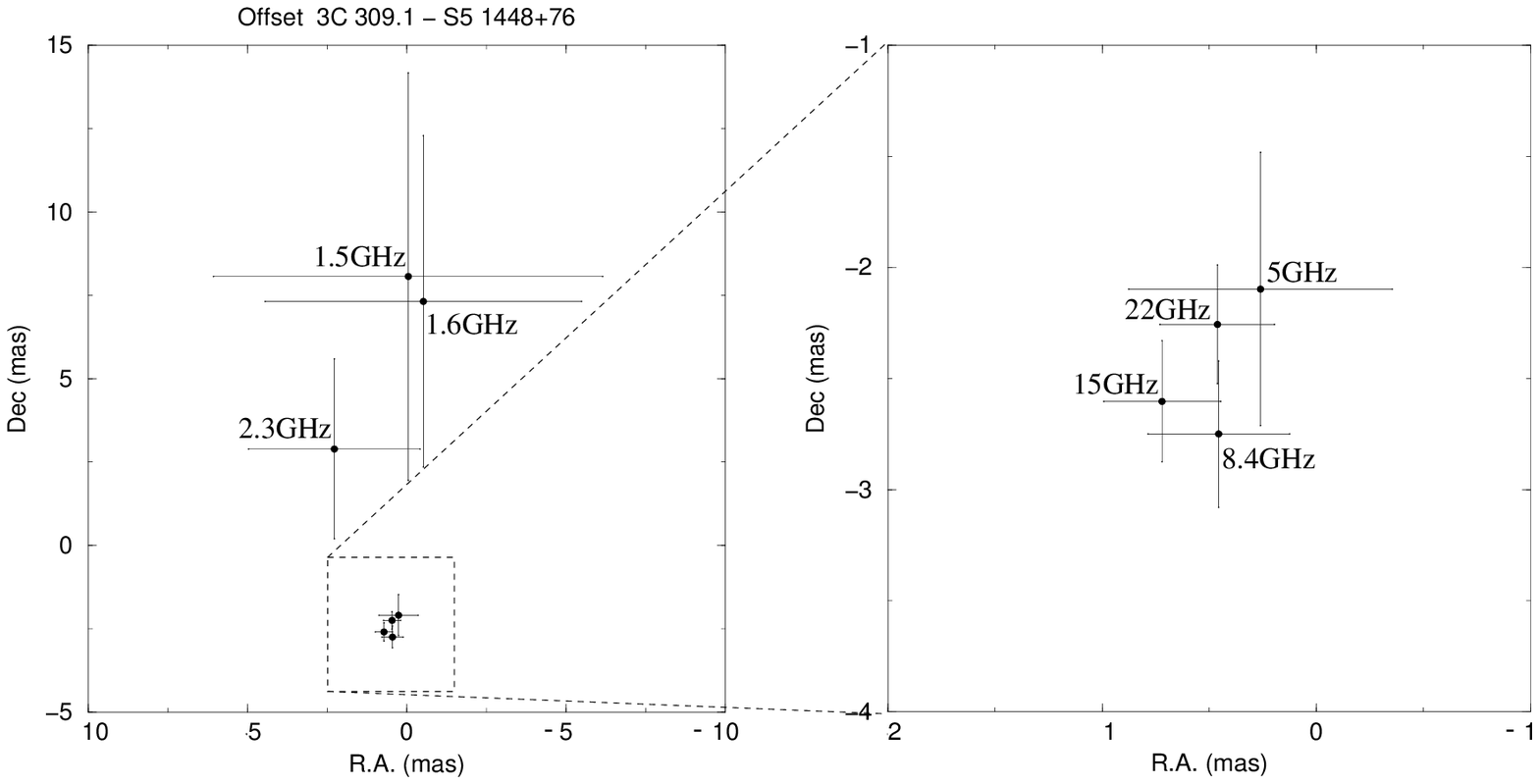}
\includegraphics{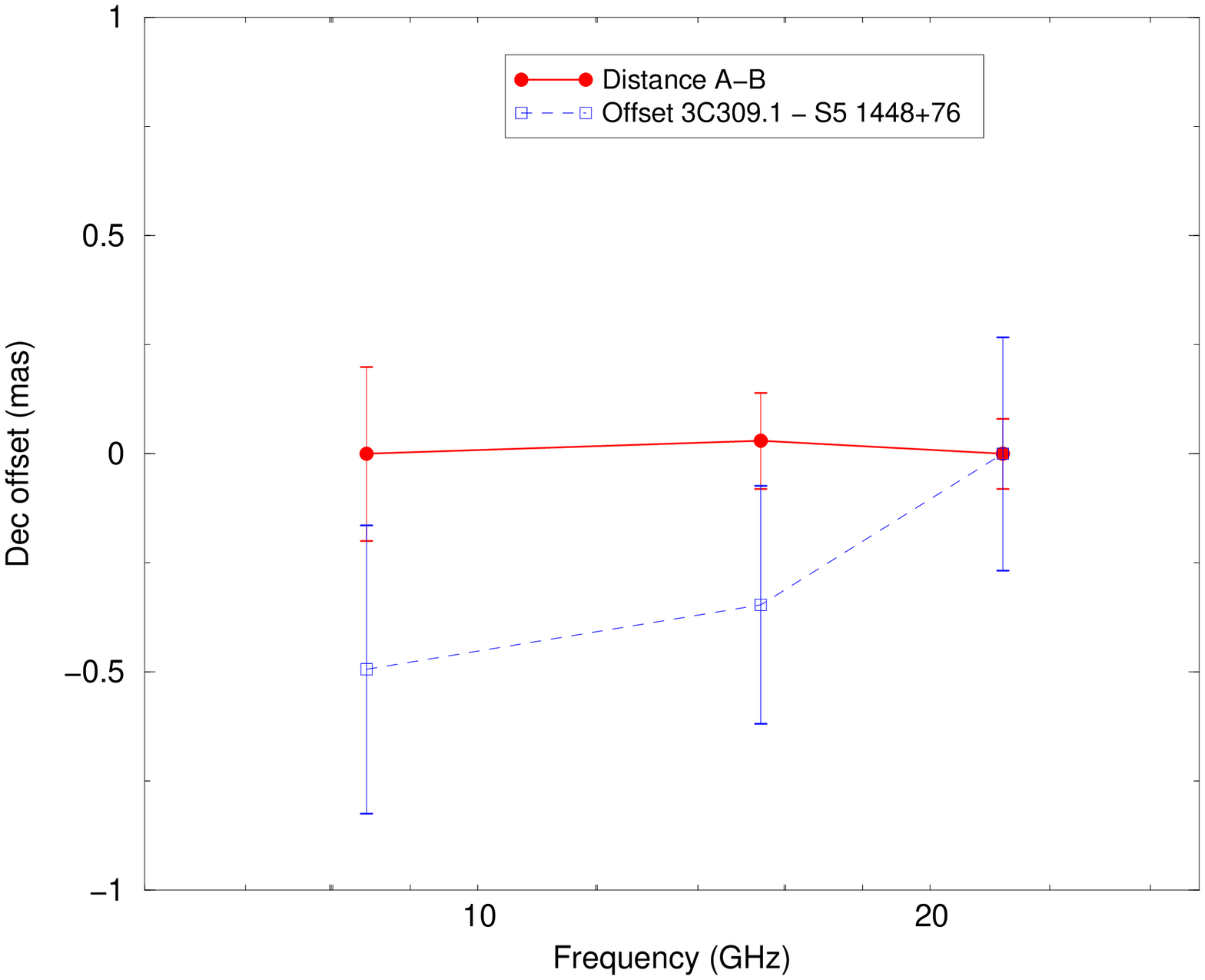}
\caption{{\bf Left and central panel:} Relative position of 
3C\,309.1 with respect
to S5\,1448+76, obtained via {\sc aips} phase-referencing as described
in the text.  The data at the lowest frequencies are dominated by
the ionospheric effect and are invalid for our purposes.  
{\bf Right panel:} Comparison between the two
different methods presented in the text to determine the core
offset at 8.4, 15 and 22\,GHz: the distance A-B with respect to the phase
referencing results.  The values at 22\,GHz are set as zero reference.
The A-B offset is biased due to beam effects in the
L-shaped B component.  The astrometric results are biased by the unmodelled
ionosphere and a simplistic model for the troposphere.
\label{fig:position1448_coreoffset}
}
\end{figure}

The error budget in the 
positions (uncertainties in Table~\ref{table:positions})
includes the following error terms:
a priori coordinates of the source, determination of peak-of-brightness in
the maps, polar motion (estimated error of 1\,mas), UT1--UTC (10$^{-4}$s), 
station coordinates (5\,cm), troposphere, ionosphere ($\propto \nu^{-2}$), 
and problems
in the {\sc aips} phase connection.  
This constitutes a conservative estimate of
the uncertainty.  We consider thus the 
phase-referencing results with S5\,1448+76 at the highest frequencies
as correct (central panel of 
Fig.~\ref{fig:position1448_coreoffset}).

\paragraph{The core position}
We assign the peak offset in $\alpha$ to S5\,1448+76 and in $\delta$
to 3C\,309.1.  The declination offsets in 3C\,309.1
between contiguous frequencies are 
shown in Table~\ref{table:coreoffset}.  The relative offsets are
also plotted in Fig.~\ref{fig:position1448_coreoffset}, where the
value at 22\,GHz has been set to be zero.

\begin{table}[htbp]
\caption{Core shift in {\bf declination} for 3C\,309.1 at the higher 
frequencies (see right panel in Fig.~\ref{fig:position1448_coreoffset}).
\label{table:coreoffset}}
\begin{center}
\begin{tabular}{crr}
                 & \mton{\small Distance A-B} & \mton{\small {\sc aips} phase-}\\
{\small Frequencies} & \mton{\small in maps} & \mton{\small referencing}\\ \hline
15\,--\,8.4\,GHz & --130$\pm$140\,$\mu$as  &  150$\pm$430\,$\mu$as \\
22\,--\,15\,GHz  &   130$\pm$140\,$\mu$as  &  350$\pm$380\,$\mu$as \\ \hline
\end{tabular}
\end{center}
\end{table}

An alternative way to measure the core offset is to assume that
the B component in the maps from Section~\ref{sec:results} is optically 
thin and its peak is at
the same position for all frequencies.  These values are also
presented in Table~\ref{table:coreoffset}, and in the right
panel of Fig.~\ref{fig:position1448_coreoffset}.

The trend in the dependence of the core position with the frequency
is different in both methods.  The A--B separation measurements
is apparently frequency-independent: the difference at
the beams at different frequencies may bias
this result, since the structure 
of the B component is L-shaped and in the declination 
coordinate it is more extended to the South.    
Our preliminary positional results from the {\sc aips}
astrometry at 8.4, 15 and 22\,GHz (but not at 5\,GHz)
suggest that the peak-of-brigthness of
the maps shifts closer to the jet basis (core at infinite
frequency) at higher frequencies, being this jet basis 
to the North of the A feature. 
This would be the expected
opacity shift produced by the synchrotron self-absorption in the jet.
Assuming that $k_r$=1 (self-absorbed core, \cite{bla79})
at 8.4\,GHz and that $R_{\rm core}\propto\nu^{-1/k_r}$, 
the astrometric results provide values of
$k_r$=1.1$\pm$0.5 at 15\,GHz and $k_r$=0.9$\pm$0.6 at 22\,GHz.  
The big uncertainties do not permit to draw any conclusions about
the physical parameters of the jet from $k_r$ at the present
status of the analysis.
A detailed analysis with the final, unbiased astrometric
results will be published elsewhere.  


\section{Summary}

We have presented preliminary results from 
a detailed multi-frequency study of the QSO
{3C\,309.1} based on the VLBA observations made in mid 1998.  
We find a curved jet extending up
to 100\,mas to the East at low frequencies with two main components,
A and B. The A component has a turnover frequency around 8.4\,GHz and
the B component is optically thin.  
The polarized intensity map at
1.5\,GHz shows that the core is un polarized.
In the region southern to B the electric field has a radial structure.  The 
external parts of the jet have a high degree of polarization. 
A preliminary astrometric analysis provides
a determination of the core position at different frequencies by
phase-referencing to a nearby radio source QSO S5\,1448+76.  
The changes at the core position with frequency suggest
high opacity close to the core caused by synchrotron self-absorption.
Due to the big uncertainties we cannot make any assert about the
value of $k_r$ at high frequencies.
An exhaustive analysis including
ionospheric and tropospheric bias removal and physical modeling of
the source will be presented in a forthcoming paper.

\vspace*{10pt}

\noindent
\begin{small}
{\bf Acknowledgements.}
We acknowledge Dr.\ Scott E.\ Aaron for his support during the data
collection and Dr.\ Richard W.\ Porcas for his critical
reading of the manuscript.
The National Radio Astronomy
Observatory is a facility of the National Science Foundation operated
under cooperative agreement by Associated Universities, Inc.
\end{small}

{}

\begin{thebibliography}{99}

\bibitem{koe81}
K\"onigl, A.\
1981, AJ 243, 700

\bibitem{lob98}
Lobanov, A. P.
1998, 
A\&A 330, 79

\bibitem{leb01}
Lebach, D. E., Ransom, R. R., Ratner, M. I., Shapiro, I. I.,
Bartel, N., Bietenholz, M. F., Lestrade, J.-F., 
2001,
in 
"Galaxies and their Constituents at the Highest Angular Resolution",
IAU Symp.\ 205, p.\ 59

\bibitem{bri00}
Brisken, W. F., Benson, J. M., Beasley, A. J., Fomalont, E. B., Goss, W. M.,
Thorsett, S. E., 
2000,
ApJ 541, 959

\bibitem{per00}
P\'erez-Torres, M. A.,  Marcaide, J. M., 
Guirado, J. C., Ros, E., Shapiro, I. I., Ratner, M. I.,
Sard\'on, E.,
2000,
A\&A 360, 161

\bibitem{ros99}
Ros, E., Marcaide, J. M., Guirado, J. C., Ratner, M. I., Shapiro, I. I., 
Krichbaum, T. P., Witzel, A., Preston, R. A.,
1999,
A\&A 348, 381

\bibitem{vbr84}
van Breugel, W., Miley, G., Heckman, T.,
1984,
AJ 89, 5

\bibitem{fan85}
Fanti, C., Fanti, R., Parma, P., Schilizzi, R. T., van Breugel, W. J. M.\
1985,
A\&A 143, 292

\bibitem{for90}
Forbes, D. A., Crawford, C. S., Fabian, A. C., Johnstone, R. M.\
1990, MNRAS 244, 680

\bibitem{aar96}
Aaron. S. E., 
1996, 
PhD Thesis, Brandeis University, MA, US

\bibitem{aar97}
Aaron, S. E., Wardle, J. F. C., Roberts, D. H.\
%
1997,
Vistas in Astronomy 41, 225

\bibitem{kus90}
Kus, A. J., Wilkinson, P. N., Pearson, T. J., Readhead, A. C. S.\
1990,
in: Parsec-Scale Radio Jets, ed.\ J.A.\ Zensus, T.J.\ Pearson,
Cambridge University Press, 161

\bibitem{she95}
Shepherd, M. C., Pearson, T. J., Taylor, G. B. 1995, BAAS 26, 987

\bibitem{lep95}
Lepp\"anen, K. J., Zensus, J. A., Diamond, P. J. 1995, AJ 100, 2479

\bibitem{bla79}
Blandford, R. D., K\"onigl, A.\
1979, ApJ 232, 34


%
%
%
%
%
%

\end{thebibliography}
\end{document}